\documentclass[prl,twocolumn,showpacs,preprintnumbers]{revtex4}
\usepackage{amssymb}

\usepackage{graphicx}

\input{tcilatex}

\begin{document}

\title{Electromechanical Effects in Carbon Nanotubes}
\author{M. Verissimo-Alves$^1$}
\email{verissim@if.ufrj.br}
\author{Belita Koiller$^1$}
\author{H. Chacham$^2$}
\author{R. B. Capaz$^1$}
\affiliation{$^1$Instituto de F\'\i sica, Universidade Federal do Rio de Janeiro, Caixa
Postal 68528, 21845-970, Rio de Janeiro, RJ, Brazil}
\affiliation{$^2$Departamento de F\'\i sica, ICEx, Universidade Federal de Minas Gerais,
Caixa Postal 702, 30123-970 Belo Horizonte, MG, Brazil}
\date{\today}

\begin{abstract}
We perform \textit{ab initio} calculations of charged graphene and
single-wall carbon nanotubes (CNTs). A wealth of electromechanical behaviors
is obtained: (1) Both nanotubes and graphene expand upon electron injection.
(2) Upon hole injection, metallic nanotubes and graphene display a
non-monotonic behavior: Upon increasing hole densities, the lattice constant
initially contracts, reaches a minimum, and then starts to expand. The hole
densities at minimum lattice constants are 0.3 $|e|$/atom for graphene and
between 0.1 and 0.3 $|e|$/atom for the metallic nanotubes studied. (3)
Semiconducting CNTs with small diameters ($d \lesssim 20$ \AA) always expand
upon hole injection; (4) Semiconducting CNTs with large diameters ($d
\gtrsim 20$ \AA) display a behavior intermediate between those of metallic
and large-gap CNTs. (5) The strain versus extra charge displays a linear
plus power-law behavior, with characteristic exponents for graphene,
metallic, and semiconducting CNTs. All these features are physically
understood within a simple tight-binding total-energy model.
\end{abstract}

\maketitle

%hc More than 10 years after their discovery by Iijima \cite{iijima} and under 
%hc intense and ever-growing research since then, carbon nanotubes (CNTs) are still 
%hc capable of unveiling surprising and fascinating physical properties. Perhaps 
%hc their most unique feature is their properties' dependence on the rolling 
%hc direction (chirality) and diameter. For instance, CNTs can be either metallic 
%hc or semiconducting depending on their chirality and diameter 
%hc \cite{hamada, saito, wildoer}.  Because of this rich variety of electronic 
%hc behaviors, many uses have been devised for CNTs in the emerging field of 
%hc nanoelectronics \cite{diodes, fets}.

%hc However, not only the electronic properties of CNTs are attractive, as these 
%hc materials have superb mechanical properties as well \cite{book_dresselhaus}, 
%hc making their use in nanomechanical applications very promising. For example, 
%hc mechanically-manipulated nanoscale bearings made of multiwall CNTs have already 
%hc been realized \cite{cumings}. The possibility of \textit{electronic} control of 
%hc nanomechanical devices would certainly bring a great improvement in their 
%hc effective switching times, control and precision \cite{forro}.   

More than 10 years after their discovery by Iijima \cite{iijima}, carbon
nanotubes (CNTs) are still capable of unveiling surprising and fascinating
physical properties. An important feature of CNTs is the dependence of their
electronic properties on diameter and chirality \cite{hamada, saito,
wildoer, diodes, fets}. However, not only the electronic properties of CNTs
are attractive, as these materials have superb mechanical properties as well %
\cite{book_dresselhaus}, making their use in nanomechanical applications
very promising \cite{cumings}. The possibility of \textit{electronic}
control of nanomechanical devices would certainly bring a great improvement
in their effective switching times, control and precision \cite{forro}.

For all these reasons, it is important to study the electromechanical
properties of CNTs. A few experimental works have focused on
electrostatically-driven mechanical responses of CNTs \cite{ugarte,lieber}.
However, mechanical response of CNTs can be driven not only by
electrostatics, but also by quantum-mechanical effects. Indeed, it is well
known that the in-plane lattice constant of intercalated graphite expands or
contracts relative to pure graphite because of 
%hc bond weakening or strengthening due to charge transfer between dopants and the graphene 
%hc planes, an intrinsically quantum-mechanical effect \cite{dress+dress, jpcs}. 
charge-transfer effects \cite{dress+dress, jpcs}. Recently, the mechanical
response of "nanotube sheets" - sheets of entangled single-wall CNTs bundles
- to electrochemical charge injection has been investigated, and its use in
actuators has been proposed \cite{baughman}. In that work, the
electromechanical response of CNTs has been found to be stronger and more
non-linear than that of graphite. Controlling and optimizing the
electromechanical response of nanotube sheets is a difficult task, 
%hc since it involves problems like bundling and entanglement of nanotubes and 
%hc disorder in the sheets. For these reasons, it would be highly desirable to 
since it involves problems like bundling and entanglement of CNTs. For these
reasons, it would be highly desirable to investigate the electromechanical
properties of \textit{individual} nanotubes and their dependence on
chirality and diameter. Since experimental control of these parameters is
still beyond the current state-of-art, the predictive power of \textit{ab
initio} theoretical calculations makes them the method of choice for such
studies.

Initially, we perform \textit{ab initio} calculations of charged graphene
and single-wall metallic and semiconducting CNTs. Our calculations are
performed within the Density Functional Theory (DFT) and pseudopotential
frameworks with a numerical-atomic-orbitals basis set (SIESTA code) \cite%
{siesta}. SIESTA has been successfully used in a number of studies on CNTs %
\cite{verissim,siesta_CNTs}.

\begin{figure}[tb]
\includegraphics[width=0.5\textwidth]{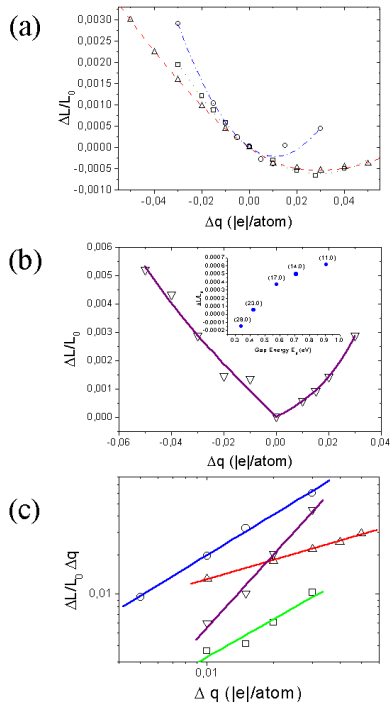} 
\caption{(a) Relative variation of the lattice constant (for graphene) or
the axial unit cell lengths (for metallic CNTs), $\frac{\protect\delta L}{L_0%
}$, as a function of the extra charge, $q$. Points are \textit{ab initio}
results, lines are best fits using Eqs. (\ref{eq:dldqgraphene}) and (\ref%
{eq:dldqmetalcnts}). (b) Same as (a) for semiconducting CNTs and fit using
Eq. (\ref{eq:dldqsemic}). Inset: plot of $\frac{\protect\delta L}{L_0}%
(q=0.01)$ for semiconducting CNTs of different diameters, showing the
crossover to "metallic-like" behavior. (c) Log-log plots showing the
different power laws of $\protect\delta L (q)$, as discussed in the text.
The slopes are $0.52\pm 0.04$, $0.94\pm 0.15$, $1.32\pm 0.21$ and $1.90\pm
0.15$, for graphene, (12,0), (5,5) and (11,0) CNTs respectively, consistent
with the predictions of the TB model ($(\protect\alpha-1)$, where $\protect%
\alpha$ is defined in Eq. (\ref{eq:simple})): 1/2, 1, 1 and 2.}
\label{fig:variosnts}
\end{figure}

Fig.~\ref{fig:variosnts}(a) shows the relative variation $\frac{\delta L}{L_0%
}$ of the in-plane lattice parameter for graphene or the unit cell axial
length \cite{footbk} for metallic zig-zag (12,0) and armchair (5,5) CNTs, as
a function of the extra charge \cite{foothelio} (in units of $|e|$) per
atom, $q$. Negative $q$ (extra electrons) cause expansion of graphene and
metallic CNTs, whereas low positive $q$ (holes) cause contraction. Large
amounts of hole injection eventually cause expansion of graphene and CNTs,
in qualitative agreement with experiments \cite{baughman}. Fig.~\ref%
{fig:variosnts}(a) also shows, from the comparison between the results for
the (5,5) and the (12,0) CNTs, that the electromechanical response of
metallic CNTs is sensitive to chirality and/or diameter. Also in agreement
with experimental observations, the strain $\frac{\delta L}{L_0}$ vs. $q$
appears to be more non-linear for CNTs than for graphene. The largest
negative strain is -0.05\% for graphene and -0.06\% for (12,0) nanotubes,
therefore larger than the experimental value of -0.02 to -0.03\%, indicating
that the effect of bundling in nanotube sheets may hinder the
electromechanical response of CNTs. A linear fit in the $q<0$ regime for
graphene gives $\frac{\delta L}{L_0q}=0.060$, in good agreement with the
experimental value of $0.066$ for intercalated graphite \cite{baughman}.

Fig.~\ref{fig:variosnts}(b) shows the behavior of $\frac{\delta L}{L_0}$ vs. 
$q$ for a semiconducting (11,0) CNT. Surprisingly, the CNT expands
regardless of the sign of $q$, in contrast to what is observed for graphene
and metallic CNTs. 

A key element to understand these results is the energy positioning of
bonding and anti-bonding states in graphene and CNTs. A widely accepted
view, based on a $\pi$-orbital, nearest-neighbor ($nn$) tight-binding (TB)
description of the energy bands of graphene \cite{book_dresselhaus}, is that
the lower (valence or $\pi$) band has a bonding character and the upper
(conduction or $\pi^*$) band has an anti-bonding character. Indeed, within
such a description, the energy bands of undistorted graphene are 
\begin{equation}
\varepsilon_0(\vec k)=\epsilon_{2p}\pm t_0 f(\vec k),  \label{eq:bsnn}
\end{equation}
where $\epsilon_{2p}$ is an ``atomic level'' for the $2p$ state of the C
atom (corrected by the crystalline potential), $t_0$ is the \textit{nn}
hopping integral for an undistorted graphene lattice (lattice constant $a_0$%
), and $f(\vec k)$ is a function of the wave vector in the 2-d Brillouin
Zone (BZ) \cite{book_dresselhaus}. In this case, $\epsilon_{2p}$ coincides
with the Fermi level, $E_F$, for a neutral system at the $K$ points of the
BZ, where $f(\vec k) = 0$. Therefore, extra electrons would occupy
anti-bonding states and extra holes would go to bonding states. In either
case the lattice would expand. Therefore, the \textit{nn} TB approximation
cannot explain the contraction of graphene and metallic CNTs for low
positive $q$.

\begin{figure}[htb]
\begin{center}
\includegraphics[width=0.25\textwidth]{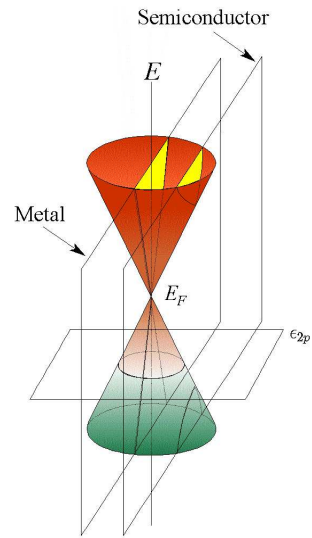}
\end{center}
\caption{(color) Schematic band structure of graphene and CNTs, in the
vicinity of a $K$ point. The green and red portions of the energy surface
indicate bonding and anti-bonding states (energy below and above $\protect%
\epsilon_{2p}$), respectively. The two vertical planes indicate the effects
of $k$-space quantization for metallic and semiconducting CNTs.}
\label{fig:bstruc4all}
\end{figure}

However, as pointed out by Kertesz \cite{kertesz}, inclusion of
next-nearest-neighbor (\textit{nnn}) interactions lifts the energy
eigenvalues at the $K$ points above $\epsilon_{2p}$, therefore making these
states antibonding. Since our method is based on an atomic-orbital basis, we
can provide \textit{ab initio} verification of these ideas. From the \textit{%
ab initio} Hamiltonian (with interactions up to $4^{th}$ \textit{nn}), we
estimate $\epsilon_{2p}$ to be roughly 1.6 eV below $E_F$ for graphene. This
is illustrated in Fig.~\ref{fig:bstruc4all}. Therefore, the states around $%
E_F$ are antibonding for both bands, and now if electrons are gradually
removed from the $\pi$ band (positive charging), the bond lengths will
contract, in agreement with experiments \cite{baughman} and with the results
in Fig.~\ref{fig:variosnts}. If positive charging continues, electrons will
be eventually removed from states below $\epsilon_{2p}$ (therefore from
bonding states) and the bond lengths will start to increase again, also in
agreement with the \textit{ab initio} results.

We explain this behavior in a more quantitative way via a total-energy TB
model. The total energy of graphene or CNTs is a function of the number of
electrons per atom, $N$, and the bond length variation $\delta L$, and it is
expressed as 
\begin{equation}
E_{tot}^N(\delta L)=E_b+E_r=2\sum_{\vec k}^{occ}\varepsilon_{\delta L} (\vec
k)+E_r  \label{eq:etottb}
\end{equation}
where $E_b$ is the band energy and $E_r$ is a repulsive energy. The
summation runs over occupied states (accounting for the proper value of $N$)
and $\varepsilon_{\delta L}(\vec k)$ is the band structure for a deformation 
$\delta L$ .

From Eq. (\ref{eq:bsnn}), we can write the band structure dependence on the
deformation as 
\begin{equation}
\varepsilon _{\delta L}(\vec{k})=\epsilon _{2p}+(\varepsilon _{0}(\vec{k}%
)-\epsilon _{2p})\frac{t_{\delta L}}{t_{0}},  \label{eq:bsdeform}
\end{equation}%
where $t_{\delta L}$ is the hopping matrix element for a distorted lattice.
Assuming a power-law dependence of $t$ on the bond length $L$, i.e. ($%
t_{\delta L}=t_{0}(L_{0}/L)^{\alpha })$ and taking the limit of small $%
\delta L$, we can linearize: 
\begin{equation}
\varepsilon _{\delta L}(\vec{k})=\varepsilon _{0}(\vec{k})-\zeta \delta
L(\varepsilon _{0}(\vec{k})-\epsilon _{2p}),  \label{eq:linearize}
\end{equation}%
where $\zeta =\alpha /{L_{0}}$. Let us consider the case of a neutral
system, with $N_{0}$ electrons. From Eqs. (\ref{eq:etottb}) and (\ref%
{eq:linearize}), $E_{b}$ is given by 
\begin{equation}
E_{b}^{N_{0}}(\delta L)=E_{b}^{N_{0}}(0)-\zeta \delta
L[E_{b}^{N_{0}}(0)-N_{0}\epsilon _{2p}].  \label{eq:ebneutral}
\end{equation}%
For a neutral system, by definition, the total energy must be minimum at $%
\delta L=0$, so 
\begin{equation}
E_{tot}^{N_{0}}(\delta L)=E_{tot}^{N_{0}}(0)+\beta (\delta L)^{2},
\label{eq:etotmin}
\end{equation}%
where $\beta $ is an elastic constant. Substitution of (\ref{eq:ebneutral})
and (\ref{eq:etotmin}) in (\ref{eq:etottb}) yields an expression for the
repulsive energy of a neutral system at arbitrary $\delta L$: 
\begin{equation}
E_{r}^{N_{0}}(\delta L)=E_{r}^{N_{0}}(0)+\beta (\delta L)^{2}+\zeta \delta
L(E_{b}^{N_{0}}(0)-N_{0}\epsilon _{2p})  \label{eq:erneutral}
\end{equation}

Consider now a system with an arbitrary number of electrons $N$. The
calculation of $E_b$ is similar to that for the neutral case, yielding 
\begin{equation}
E_b^N(\delta L)=E_b^N(0)-\zeta\delta L(E_b^N(0)-N\epsilon_{2p}).
\label{eq:ebcharged}
\end{equation}
We also assume that, to first order, the dependence of $E_r$ on the bond
length is not very much affected by the extra charge. Then, from Eq. (\ref%
{eq:erneutral}) we have 
\begin{equation}
E_r^N(\delta L)=E_r^N(0)+\beta(\delta L)^2+\zeta\delta L(E_b^{N_0}(0)-N_0
\epsilon_{2p})  \label{eq:ercharged}
\end{equation}

Combining Eqs. (\ref{eq:etottb}), (\ref{eq:ebcharged}) and (\ref%
{eq:ercharged}), and defining the extra charge as $q=-\Delta N=N_0-N$, we
finally arrive at an expression for the total energy of a charged system: 
\begin{equation}
E_{tot}^{q}(\delta L)=E_{tot}^{q}(0)+\beta(\delta L)^2+\zeta\delta L
q(\Delta\tilde E_b-\epsilon_{2p}).  \label{eq:etotcharged}
\end{equation}
Here, $\Delta \tilde E_b=\frac{E_b^N(0)-E_b^{N_0}(0)}{\Delta N}$ is the
variation in band energy per extra electron. By imposing $\frac{\partial
E_{tot}}{\partial L}|_{q}=0$, we obtain the unit cell length distortion as a
function of $q$: 
\begin{equation}
\delta L(q)=-\frac{\zeta}{2\beta}q\left(\Delta \tilde
E_b-\epsilon_{2p}\right).  \label{eq:dldq}
\end{equation}

Eq. (\ref{eq:dldq}) has a very simple physical meaning. Consider for
instance electron injection ($q < 0$). If $\Delta \tilde E_b-\epsilon_{2p} >
0$, extra electrons are predominantly placed in anti-bonding levels, so the
lattice expands ($\delta L >0$). Similar arguments can be used for electrons
in bonding levels (yielding lattice contraction) and holes in anti-bonding
or bonding levels (yielding contraction or expansion).

We now calculate $\delta L(q)$ for graphene and CNTs. Let us consider
graphene first. In the vicinity of $E_F$, we can use a simple ``conical''
approximation for the energy bands, $\varepsilon(\vec k)=E_F\pm\gamma|\vec k|
$, where $+$ stands for electrons and $-$ for holes, and the origin in $k$%
-space is shifted to the $K$ point. $\Delta \tilde E_b$ may then be obtained
by integrating $\varepsilon(\vec k)$ over the region of the BZ occupied with
extra electrons, i.e. a disk of radius $\Delta k$ which is related to $q$ as 
$\mp q=\frac{s(\Delta k)^2}{\pi}$, where $s$ is the area per atom. After
integrating and substituting the result into Eq. (\ref{eq:dldq}), we get 
\begin{equation}
\delta L(q)=-\frac{\zeta}{2\beta}q\left[E_F- \epsilon_{2p}\pm\frac{2\gamma}{3%
}\left(\frac{\pi |q|}{s}\right)^ \frac{1}{2}\right].  \label{eq:dldqgraphene}
\end{equation}

The energy bands of CNTs are obtained by slicing the energy bands of
graphene along the proper quantization lines \cite{book_dresselhaus}. We
show schematically in Fig.~\ref{fig:bstruc4all} the result of this procedure
in the vicinity of one of the $K$ points. For metallic CNTs, one of the
quantization lines will cross the $K$ point, yielding linear 1-d bands. If
the CNT is semiconducting, the 1-d bands will be parabolic.

For metallic CNTs, the calculation of $\Delta \tilde E_b$ is very similar,
the only essential difference being that the integration is now
one-dimensional and the integration limits are related to the $q$ as $\mp q=%
\frac{\Delta k \ell}{\pi}$, where $\ell$ is the CNT length per atom. The
result is 
\begin{equation}
\delta L\left(q\right)=-\frac{\zeta}{2\beta}q\left(E_F- \epsilon_{2p}-\frac{%
\gamma\pi q}{2\ell}\right).  \label{eq:dldqmetalcnts}
\end{equation}

Finally, for semiconducting CNTs we use a parabolic dispersion $\varepsilon
(k)=E_F\pm\frac{\varepsilon_g}{2}\pm\eta k^2$, where ${\varepsilon_g}$ is
the energy gap. The result for $\delta L(q)$ is 
\begin{equation}
\delta L(q)=-\frac{\zeta}{2\beta}q\left[E_F\pm \frac{\varepsilon_g}{2}%
-\epsilon_{2p}\pm\frac{\eta\pi^2}{12\ell^2}q^2 \right].  \label{eq:dldqsemic}
\end{equation}

From Eqs. (\ref{eq:dldqgraphene}), (\ref{eq:dldqmetalcnts}) and (\ref%
{eq:dldqsemic}) we see that the electromechanical behaviors of graphene and
CNTs can all be described by simple expressions: 
\begin{equation}
\frac{\delta L(q)}{L_0}=-aq + b|q|^\alpha ,  \label{eq:simple}
\end{equation}
where the linear coefficient $a$ controls the electromechanical response for
low charge injection and depends precisely on the positioning of $E_F$ with
respect to $\epsilon_{2p}$. The non-linear term is defined by the exponent $%
\alpha$, which is 3/2, 2 and 3 for graphene, metallic and semiconducting
CNTs. We test these predictions from the \textit{ab initio} results. In Fig.~%
\ref{fig:variosnts}(c) we display log-log plots of $\frac{\delta L}{L_0 q} +
a$ vs. $|q|$ for each system. Within error bars, the exponents predicted by
the simple TB model are beautifully confirmed by the \textit{ab initio}
results. Notice that both (5,5) and (12,0) metallic CNTs have the same
exponent $\alpha=2$. The stronger non-linearity in strain vs. $q$ for the
(5,5) CNT - evident from Fig.~\ref{fig:variosnts}(a) - is due to a higher
coefficient of the quadratic term in $\frac{\delta L}{L_0}(q)$. Such a large
difference in the quadratic term indicates a strong dependence of the
non-linear response of metallic CNTs on their chirality.

Let us consider again semiconducting CNTs. From Fig.~\ref{fig:bstruc4all}
and Eq. (\ref{eq:dldqsemic}), we see that, for large enough $\varepsilon _g$%
, the upper valence states will stay below $\epsilon_{2p}$. Therefore, for
positive $q$, electrons will be removed from bonding states and expansion
will occur. This is precisely what Fig.~\ref{fig:variosnts}(b) shows. On the
other hand, we can predict that if $\varepsilon _g$ is small enough,
semiconducting CNTs will behave qualitatively like metallic ones, i.e.,
contracting upon hole injection. Since, $\varepsilon_ g \propto 1/d^2$
(where $d$ is the CNT diameter) \cite{book_dresselhaus}, we test this
prediction by performing \textit{ab initio} calculations on zig-zag
semiconducting CNTs of different diameters. The inset of Fig.~\ref%
{fig:variosnts}(b) shows a plot of $\delta L/L_0$ as a function of $d$ for a
fixed, small extra charge of 0.01 holes/atom. This indeed shows that the
decrease in the gap of $(n,0)$ semiconducting CNTs with increasing $n$
ultimately leads to a recovery of metallic nanotube behavior.

In conclusion, we have performed \textit{ab initio} studies of
electromechanical effects in graphene and CNTs. The diversity on
electromechanical responses is fascinating, and it has an amazingly simple
explanation in terms of the bonding and anti-bonding nature of the
electronic states within a TB model. The observed stronger non-linear
behavior in $\delta L(q)$ for CNTs as compared to graphene is reproduced and
explained, arising naturally as a dimensionality effect in the distribution
of extra electrons over the BZ's of graphene and CNTs. Our results provide
important guidelines for fabrication of nano-electromechanical devices.
Indeed, when experimental control of production and assembling of nanotubes
with specific chiralities is achieved, the possibilities in the architecture
of different devices by combining the variety of effects described here will
be only limited by our imagination.

\begin{acknowledgments}
This work was partially supported by Brazilian agencies CAPES, CNPq, FUJB,
FAPERJ, PRONEX-MCT and Instituto do Mil\^enio de Nanoci\^encias. We wish to
thank Maur\'{\i}cio Ara\'ujo for help with preparation of Fig.~\ref%
{fig:bstruc4all}.
\end{acknowledgments}

\textit{Note added:} After the submission of this work, a related study by
Gartstein \textit{et al} \cite{gartstein} has appeared in the literature.


\begin{thebibliography}{99}
\bibitem{footbk} Our calculations include full axial and radial relaxation
induced by charge injection. We show here only the axial changes, which are
the most relevant ones for possible applications of these electromechanical
effects. Anisotropies between radial and axial behavior, as previously
obtained \cite{verissim}, are also present here.
\bibitem{foothelio} We consider a range of injected charge , -0.05$<q<$0.05
, well within the reported experimental values for graphite intercalated
compounds \cite{baughman}. Those can be taken as estimated bounds for charge
injection in CNTs.
\end{thebibliography}
\end{document}